\documentclass[%
 reprint,
showpacs,preprintnumbers,
 amsmath,amssymb,
 aps,
prc,
superscriptaddress]{revtex4-1}

\usepackage{graphicx}
\usepackage{dcolumn}
\usepackage{bm}
\usepackage{amsmath}

\begin{document}


\title{Modified  MIT Bag Models - part II: \\ QCD phase diagram and hot quark stars}

\author{Luiz L. Lopes}
\email{luiz\_kiske@yahoo.com.br}
\affiliation{Centro Federal de Educa\c c\~ao  Tecnol\'ogica de
  Minas Gerais Campus VIII, CEP 37.022-560, Varginha, MG, Brasil}
\author{Carline Biesdorf}
\author{K. D. Marquez}
\author{D\'ebora P. Menezes}
\affiliation{%
 Departamento de Fisica, CFM - Universidade Federal de Santa Catarina;  C.P. 476, CEP 88.040-900, Florian\'opolis, SC, Brasil 
}%


\begin{abstract}

 In the present work we use the modified versions of the MIT bag model, on which both a vector field and a self-interacting term are introduced, to obtain hot quark matter and to investigate the QCD phase diagram. We first analyze two-flavored quark matter constrained to both the freeze-out and the liquid-gas phase transition at the hadronic phase. Later, three-flavored quark matter subject to $\beta$ equilibrium and charge neutrality is used to compute quark star macroscopic properties, which are confronted with recent observational massive and canonical star radius results. Finally, a comparison with QCD phase diagrams obtained from the Nambu-Jona-Lasinio model is performed.

\end{abstract}


\maketitle

\section{Introduction}

One of the most important features of the quantum chromodynamics (QCD)
is the asymptotic freedom, which predicts that strongly interacting
matter  undergoes a phase transition  from hadrons (constituted of
confined quarks) to deconfined quarks and gluons - the quark gluon
plasma (QGP) -  at some high temperature as well as high density
~\cite{Fukushima,Gross}. Therefore, the correlation
between two external parameters arises: for a fixed  temperature, what is the  density (or equivalently, the baryon chemical potential), at which the phase transition occurs in equilibrium QCD?

To study the QGP as well as the hadron-quark phase transition several experiments have been proposed and performed in recent years at LHC, RICH and
 others~\cite{Agakishiev,Adam,Abelev1,Abelev2}. Studying Au+Au and
 Pb+Pb collisions, Cleymans ~\cite{Cleymans} was able to trace
 the chemical freeze-out line, obtained when inelastic collisions between particles cease such that the abundance ratios do not change anymore~\cite{Mun}. Although the chemical freeze-out is not directly related
to the hadron-quark phase transition, the hadron multiplicities in central high energy nucleus-nucleus collisions are established very close to the phase boundary
 between hadronic and quark matter~\cite{Mun2}, especially at very low baryon chemical potential, when the chemical freeze-out temperature and
the critical temperature are expected to lie at the same error
bar~\cite{Mun2,Kar}. 
Moreover, the chemical freeze-out is expected to be a pure hadronic process, therefore,
its trace needs to be in the hadron phase. This feature acts as a
constraint for hadron-quark phase transition modeling.

In the standard model, the tool to describe strong interacting matter is the QCD. For low chemical potential and high temperature, the lattice QCD (LQCD) can be employed yielding satisfactory results.
For instance, LQCD predicts the existence of a smooth crossover
around a temperature of 160 MeV at low chemical potentials, while at higher densities a
first order phase transition~\cite{Aoki,Bell,Luo} is generally
obtained from effective models. Furthermore, the first order phase
transition must end at an unique point where a second order phase transition
takes place, the critical end point (CEP), although its existence and exact location are not well-established~\cite{Bazavov1,Bazavov2}.
Another important region in the QCD phase diagram is the liquid-gas
instability region related to nuclear fragmentation~\cite{Finn,Agostino,Debora2001}.
 For low temperatures and chemical potential, the nucleons are
 confined into the nuclei~\cite{Motornenko}, which can be regarded as a
 liquid phase. As the temperature increases,  the nuclei start to
 dissolve into a diluted interacting gas of nucleons. A critical
 temperature, above which only the gas phase survives, is expected~\cite{Elliott}.
Like the chemical freeze-out, the liquid-gas phase transition is a pure hadronic process, where the contribution from the quark degrees of freedom  can be neglected. Nevertheless, the region expected to undergo this phase transition can also be used as a constraint.     

In this work we use the modified versions of the MIT bag model, as
originally introduced in ref.~\cite{Carline} - on which  a vector
field is added in a minimal coupling scheme, 
as well as a self-interacting term that mimics the contribution from
the quark Dirac sea~\cite{Dirac} - to study the QCD phase diagram and 
hot quark matter. We start considering symmetric two-flavored quark matter,
$\mu_d =  \mu_u$, and check if we can fit both the freeze-out and the
liquid-gas phase transition at the hadronic phase. Then, we verify if
it is possible to fulfill these constraints alongside the existence of
stable strange quark matter (SQM) as proposed by the Bodmer-Witten
conjecture~\cite{Bod,Witten}. If this is true, therefore, the nuclear
matter as we know, made of protons and neutrons
is only meta-stable, and the true ground state of all matter are not
the baryons but three-flavored deconfined quark matter ($\mu_d = \mu_u = \mu_s$).

The next step is to construct a QCD phase diagram for three-flavored quark matter in $\beta$ equilibrium and zero electrical charge using the modified versions of the MIT bag model. At $T=0$, this study is important if one wants to describe 
quark (or strange) stars, as well as quark matter in
the core of massive hybrid stars \cite{Nature2020}.
In the case where the strange quark is present, there are two possibilities for the strength of its interaction with the vector field: an universal coupling, when the strength of the $s$ quark is equal
to the $u$ and $d$ quarks, and one that we calculated from the group theory approach that fixes the $s$ quark coupling constant to 2/5 of the $u$ and $d$ quarks~\cite{Carline}.
In all cases, we impose that $\beta$ stable three-flavored quark
matter needs to reproduce massive pulsars like MSP J070+6620~\cite{Cromartie}.
Another important constraint is the radius of the canonical
1.4 M$_\odot$ neutron star. A recent study using multi-messenger observations
indicates that this radius
is in the range of 10.4 km to 11.9 km~\cite{Capano}, although
less restrictive constraints using X-ray telescopes  have not been discarded~\cite{Riley,Miller}.

However, it must be clear that although the results coming from ref.~\cite{Riley,Miller,Capano} are observational constraints,
they still use some nuclear model to fit the data. Therefore the limit value of 11.9 km must be faced with care.

As will become clear in the manuscript, for reasonable values of the bag pressure parameter,
the MIT bag model and its extensions are not suitable to describe
quark matter at low chemical potential and high temperature, once its  
critical temperature is far below those (pseudo) temperatures
predicted by LQCD~\cite{Aoki,Bazavov1,Bazavov2},
as well as the experimental line of the chemical
freeze-out~\cite{Cleymans}.
 Due to these facts, inspired by an old recipe \cite{Muller,Dey}, we propose a
 simple parametrization for the Bag,
where it increases with temperature. With this simple modification we are
able to produce massive quark stars and ensure that the freeze-out and liquid-gas phase transition happen in the confined phase. At the end we study the effects of
finite temperature on hot quark stars (with a fixed temperature of 40 MeV)
and on the speed of the sound of the quark matter. Finite temperature is
important at the early stages of a quark star and in the quark core of a massive
proto-hybrid star. A recent study suggests that the presence of quark matter is linked to the behaviour of the speed of sound in massive hybrid stars~\cite{Nature2020}. 

It is also worth pointing out that in the limit of vanishing quark
masses the QCD Lagrangian presents chiral
symmetry~\cite{Buballa2005}. Although at low energy scale this symmetry is dynamically broken, it is restored at high energy scales. Besides, for low temperature and high chemical
potentials, the existence of a color superconducting phase is 
possible. Some studies indicate that the related gaps in
the fermion spectrum could be of the order of 100 MeV~\cite{Alford}.  Both features are beyond the QCD characteristics that one can study using the MIT-like models presented in this work, but for the sake of comparison, we add a section where results obtained with the Nambu--Jona-Lasinio model \cite{Nambu} are shown and commented. 

\section{Formalism} 

The MIT bag model considers that each baryon is composed of three
non-interacting quarks inside a bag. The bag, in turn, corresponds to
an infinity potential, which confines the quarks. In this simple model 
the quarks are free inside the bag and are forbidden to reach out. All
the information about the strong force lies in the bag constant, also called
the vacuum pressure. The MIT Lagrangian density 
reads~\cite{MITL}:

\begin{equation}
\mathcal{L} = \sum_{u,d,s}\{ \bar{\psi}_q  [ i\gamma^{\mu} \partial_\mu - m_q ]\psi_q - B \}\Theta(\bar{\psi}_q\psi_q), \label{e1}
\end{equation}   
where $m_q$ is the $q$ quark mass  running from $u$, $d$ and $s$, whose values are 4 MeV, 4 MeV and 95 MeV respectively~\cite{smass};
 $\psi_q$ is the Dirac quark field, $B$ is the constant vacuum pressure and $\Theta(\bar{\psi}_q\psi_q)$ is
the Heaviside step function that is included to assure that the quarks exist only confined inside the bag. 

As in ref.~\cite{Carline}, we introduce a quark interaction via minimal
coupling described by a vector channel $V_\mu$ analogous to the
$\omega$ meson in quantum hadrodynamics:

\begin{equation}
\mathcal{L}_V = \sum_{u,d,s}g_{qqV}\{ \bar{\psi}_q  [ \gamma^{\mu} V_\mu  ]\psi_q \}\Theta(\bar{\psi}_q\psi_q), \label{v0}
\end{equation}
as well as the mass term and a self-interaction on the vector field:

\begin{equation}
\mathcal{L}_V = \frac{1}{2}m_V^2V_\mu V^\mu + b_4 \frac{(g^2 V_\mu V^\mu)^2}{4} \label{v2}
\end{equation}
where $g_{qqV}$ is the coupling constant of the quark $q$ with the
vector field $V^\mu$. As pointed out earlier, there are two 
possibilities: an universal coupling with $g_{ssV} = g_{uuV} =
g_{ddV}$, as well as a ratio  that comes from 
symmetry group calculations~\cite{Carline}: $g_{ssV} = 2/5 \cdot g_{uuV} = 2/5 \cdot g_{ddV}$;
$m_V$ is the mass of the vector field, taken to be 780 MeV, $b_4$ is a
dimensionless parameter to modulate the self-interaction of the vector
field, and $g = g_{uuV}$ for short.

Now, assuming mean field approximation (MFA) ($V^{\mu} \rightarrow \langle V \rangle \rightarrow \delta_{0,\mu} V^{0}$),
we obtain the eigenvalue for the energy of the quarks and the equation of motion for the $V$ field, respectively:

\begin{eqnarray}
E_q  = \sqrt{m_q^2 + k^2} + g_{qqV}V^0, \nonumber \\
gV_0 + \bigg (\frac{g}{m_V} \bigg )^2 \bigg ( b_4(gV_0)^3 \bigg ) =
\bigg ( \frac{g}{m_V} \bigg )   \sum_{u,d,s} \bigg ( \frac{g_{qqV}}{m_V} \bigg ) n_q  \nonumber \\ \label{v3}
\end{eqnarray}
where the term $\langle\bar{\psi}_q\gamma^0\psi_q\rangle$ was replaced by  the number density $n_q$  for the $q$ quark.

Now, quarks are Fermions with spin 1/2, then the number density, the pressure  and the energy density of the quark matter can be obtained via Fermi-Dirac distribution~\cite{Greiner}:

\begin{equation}
n_q = \langle\bar{\psi}_q\gamma^0\psi_q\rangle =  2 \times N_c \int \frac{d^3k}{(2\pi)^3} (f_{q+} - f_{q-}) , \label{nd}
\end{equation} 

\begin{equation}
\epsilon_q =  2 \times N_c \int \frac{d^3k}{(2\pi)^3} E_q (f_{q+} + f_{q-}) , \label{ed}
\end{equation}

\begin{equation}
p_q =   \frac{2  \times N_c}{3} \int \frac{d^3k}{(2\pi)^3} k \frac{\partial E_q}{\partial k} (f_{q+} + f_{q-}) , \label{pres}
\end{equation}
where $N_c$  = 3  is the number of colors. The Bag contribution, as well as the contribution of the vector field  to the 
energy density and the pressure  are easily obtained through the Hamiltonian:

\begin{eqnarray}
\mathcal{H} =  -\langle \mathcal{L} \rangle = B - \frac{1}{2}m_V^2 V_0^2 
- b_4  \frac{(gV_0)^4}{4},
\end{eqnarray}
and 
\begin{equation}
\epsilon  = \sum_q \epsilon_q + \mathcal{H} . \quad \mbox{and} \quad  p = \sum_q p_q - \mathcal{H}   \label{e5}
\end{equation}
with $f_{q+}~(f_{q-})$ being the Fermi-Dirac distribution of the quarks (anti-quarks), given by:
\begin{equation}
f_{q \pm}  = \frac{1}{ 1 + \exp[(E_q^{*} \mp \mu^{*}_q)/T]}.
\end{equation}
Here, $E_q^{*} = \sqrt{k^2 + m_q^2}$ and $\mu^{*}_q =  \mu_q - g_{qqV}V_0$ is the effective chemical potential.

\section{Phase diagram and quark stars}

We start by studying how the vector field as well as the self-interacting term influence the QCD phase diagram obtained with the modified MIT model for different bag pressure values. In MIT-based models
the interpretation of the QCD phase diagram is simple: for low values of temperature and chemical potential
we have confined quarks, indicated by a negative pressure due to the bag $B$. This is the hadronic phase.
When matter is heated, light hadrons, preferentially pions, are created thermally, which increasingly fill
the space between the nucleons. Because of their finite spatial extent, the pions and other thermally produced
hadrons begin to overlap with each other and with the bags of the original nucleons such that a network of
zones with quarks, anti-quarks, and gluons is formed. At a certain critical temperature $T_c$ these zones
 fill the entire volume in a “percolation” transition. This new state of matter is the quark-gluon plasma (QGP)~\cite{Mun}.
A similar picture emerges when matter is strongly compressed. In this case the nucleons overlap at a critical number density $n_c$
and form a cold degenerate QGP consisting mostly of quarks.
Therefore, in our work the transition line between confined quark matter and QGP is indicated by $p = 0$.  
 It is important to point out that, although in our MIT-based model
 we  always treat it as a first order phase transition, at low chemical potential the LQCD predicts a smooth crossover.
 
 \begin{figure}[ht] 
\begin{centering}
 \includegraphics[angle=270,
width=0.49\textwidth]{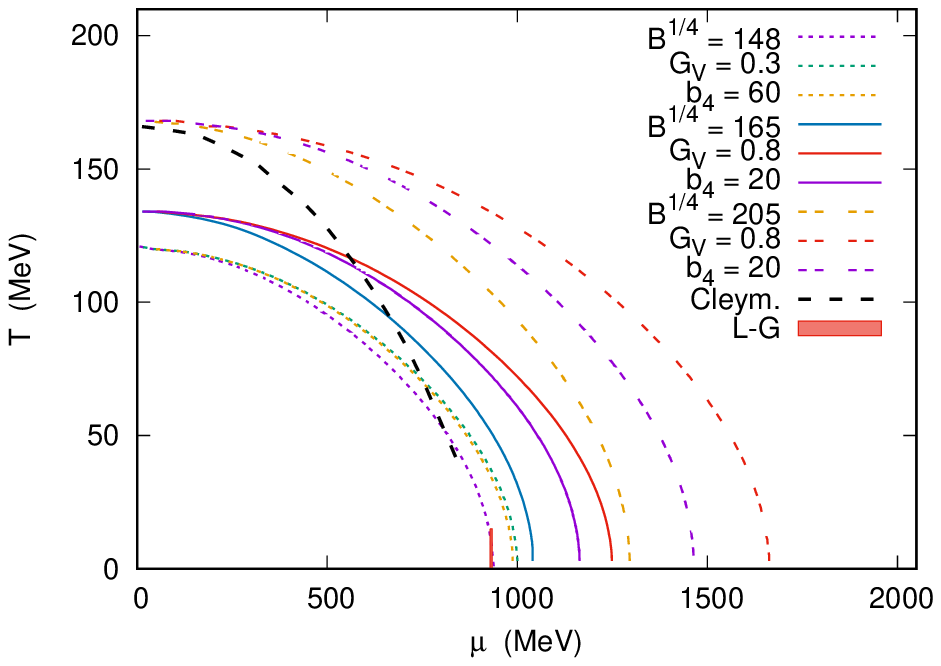}
\caption{(Color online) QCD phase diagram for different values of $B$, $G_V$ and $b_4$. The Cleym. line is the experimental freeze-out~\cite{Cleymans} and L-G is the region where we expect a liquid-gas phase transition~\cite{Finn}.}\label{F1}
\end{centering}
\end{figure}

Before we proceed, we introduce some definitions. As in our work $g_{uuV}$ and $g_{ddV}$ are always equal (and called only $g$ for short), we define
$X_V$ as the ratio between $g_{ssV}$ and $g_{uuV}$. Also, the strength of the vector channel is directly related
to $(g/m_V)^2$, so we define $G_V$ as this quantity:
\begin{equation}
X_V \doteq  \frac{g_{ssV}}{g_{uuV}} \quad \mbox{and} \quad G_V  \doteq \bigg ( \frac{g}{m_V} \bigg )^2. \label{definition}
\end{equation}

 \begin{figure*}[htb]
\begin{tabular}{cc}
\includegraphics[angle=270,width=0.49\textwidth]{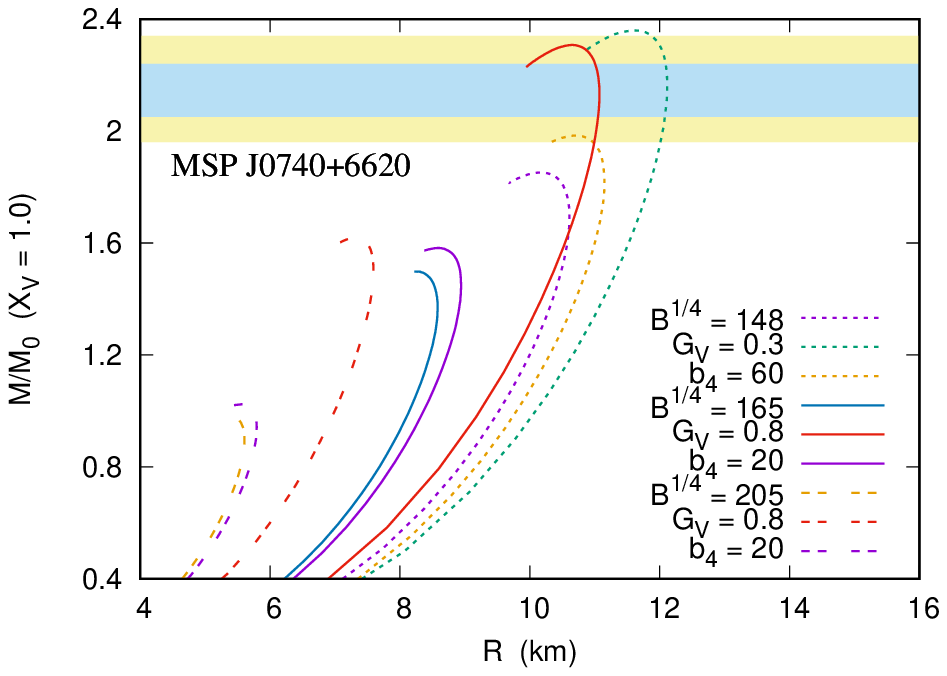} &
\includegraphics[angle=270,width=0.49\textwidth]{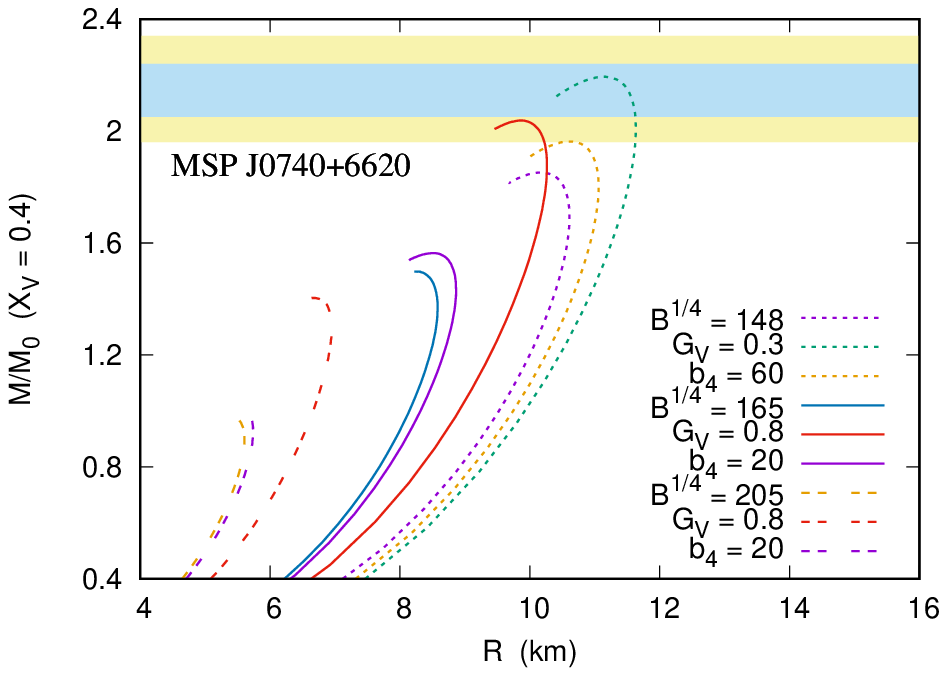} \\
\end{tabular}
\caption{(Color online) Mass-radius relation for different values of $B^{1/4}$, $G_V$ and $b_4$ with $X_V = 1.0$ (left) and $X_V = 0.4$ (right).
The hatched areas correspond to 68$\%$  and 95$\%$ credibility interval for the MSP J0740+6620.} \label{F2}
\end{figure*}
\begin{figure*}[htb]
\begin{tabular}{cc}
\includegraphics[angle=270,width=0.49\textwidth]{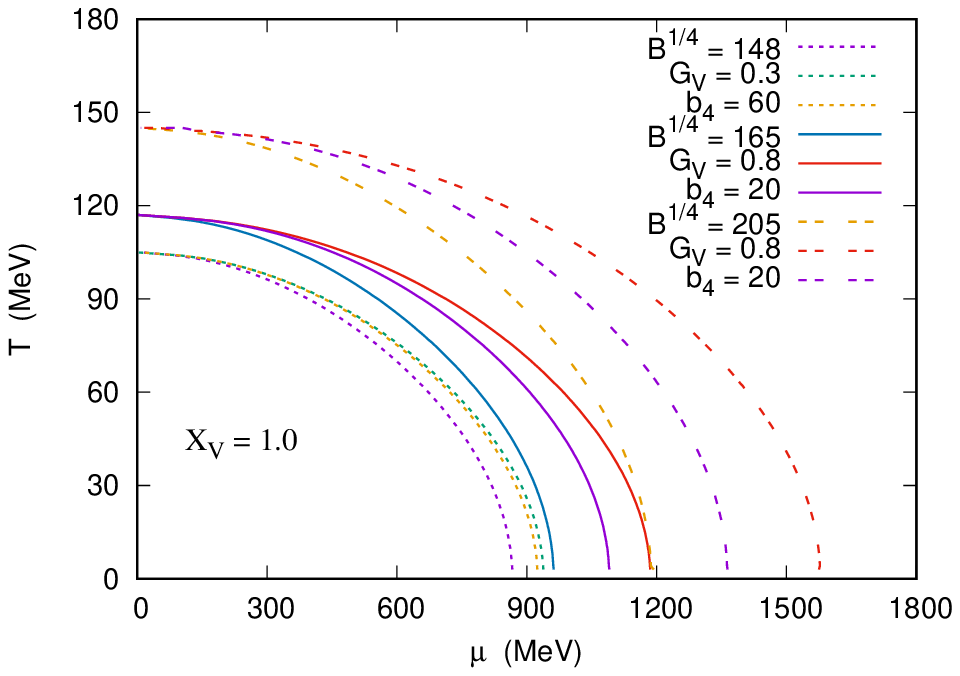} &
\includegraphics[angle=270,width=0.49\textwidth]{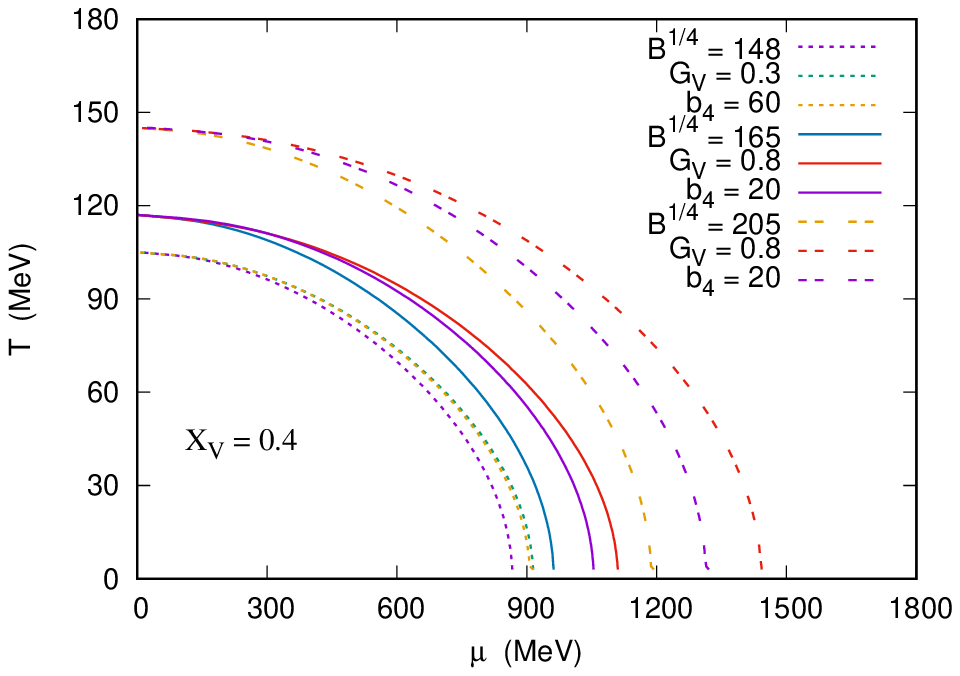} \\
\end{tabular}
\caption{(Color online) Phase diagram for $X_V$ = 1.0 (left) and $X_V$ = 0.4 (right) for neutral beta stable matter. } \label{F3}
\end{figure*}

We start with two-flavored symmetric matter with $\mu_u=\mu_d$ to make a direct comparison with the chemical freeze-out as presented in ref.~\cite{Cleymans}, as well as with
the liquid-gas phase transition. 
We use three different values for the bag parameter, $B^{1/4}$ = 148 MeV, 165 MeV and 205 MeV. $B^{1/4}$ = 148 MeV is the lowest allowed value for non-interacting bag model because even lower values would predict stable $u-d$ matter. 
A good model to simulate the hadron-quark phase transition needs to ensure that both the chemical freeze-out and the liquid-gas phase transition lie in the confined (hadronic) phase. At $T = 0$ we also expect that the phase transition takes place at values of the chemical potential at least higher than $\mu=1050$ MeV, as shown in ref.~\cite{Fukushima} using the Polyakov loop formalism. Also, using NJL-based models, the works of ref.~\cite{Buballa2005,Ruster} indicate that the critical chemical potential is around 1080 MeV to 1100 MeV. There is no experimental evidence of the maximum chemical 
potential which preserves the hadronic phase, however, a recent study points out that quark matter inside massive neutron stars is not only possible
but probable~\cite{Nature2020}. Using chiral bag model with B$^{1/4}$ = 170 MeV, ref.~\cite{Klahn3} found a critical chemical potential around  1250 MeV. For $\beta$-stable matter, using  relativistic density functional approach, ref.~\cite{Ayriyan} indicates that it should occur around $\mu=1200$ MeV; for two-flavored symmetric matter we assume
a maximum value of $\mu=1400$ MeV as a more conservative estimate.   Also, there are strong evidences that
 the (pseudo) critical temperature obtained from LQCD is very close to the critical temperature of chemical freeze-out~\cite{Mun3}. Once these constraints are investigated, we proceed by checking whether the model is able to reproduce massive quark stars, and if the parameters are in favor or against the  Bodmer-Witten conjecture of stable strange matter, i.e., the energy per baryon of strange matter $\mu_u = \mu_d = \mu_s$ must be below 930 MeV~\cite{Bod,Witten,Carline}.

Moreover, for each bag pressure value there are also three possibilities: the original MIT bag model without any interaction, a vector channel via minimum coupling and a vector channel
with minimal coupling alongside a self-interacting term to simulate the quark Dirac sea contribution. In order to reproduce massive quark stars, we use 
here $G_V = 0.3$ fm$^2$ for a bag pressure value of $B^{1/4}$ = 148 MeV, and $b_4 = 60$. For $B^{1/4}$ = 165 MeV and 205 MeV, we use $G_V = 0.8$ fm$^2$
and $b_4 = 20$. Notice that the self-interacting term depends on both, $b_4$ and $g$ as shown in Eq.~(\ref{v2}).
The results are presented in Fig.~\ref{F1}.  We also use these values and construct the 
equation of state (EoS) of a $\beta$-stable matter with zero electrical charge, namely~\cite{Rhabi}:
\begin{eqnarray}
\mu_d = \mu_s = \mu_u + \mu_e \quad \mbox{and} \quad \mu_e = \mu_\mu \nonumber \\
n_e + n_\mu = \frac{1}{3}(2n_u - n_d - n_s) . \label{cheq}
\end{eqnarray}
We then use the EoS at zero temperature as an input to the TOV equations~\cite{TOV} to produce the mass-radius diagram. The results for different values of $X_V$ are presented in Fig.~\ref{F2}, as well as
 the  MSP J070+6620 pulsar,  whose mass range is $2.14^{+0.10}_{-0.09}M_\odot$ at 68 \% credibility interval (light blue ) and
 2.14$^{+0.20}_{-0.18}M_{\odot}$ at 95 \% credibility interval (light yellow) ~\cite{Cromartie}.
We also display in Fig.~\ref{F3} the phase diagrams for three-flavored $\beta$-stable matter. Although zero temperature is generally assumed for 
calculations of stable neutron and quark stars, higher temperatures can be
important in the early stages of the proto-star. For instance, the
neutron matter deconfinement occurs on a 
strong interaction time scale of $10^{-23}$ s~\cite{Benvenuto},
while the chemical equilibrium happens 
at a weak time scale of $10^{-8}$ s. On the other hand, the cooling of the newborn
 neutron star by neutrino diffusion takes a few seconds~\cite{Lattimer}.
Therefore, the phase diagram for values up to $T=50$ MeV in neutral $\beta$-stable matter is important \cite{Debora2004}. For higher temperatures the discussion is more pedagogical, once we do not expect 
$\beta$-stable matter in this region~\cite{Gupta,Rafa2008}.

\begin{widetext}
\begin{center}
\begin{table}[ht]
\centering
\resizebox{\textwidth}{!}{%
\begin{tabular}{|c|c|c|c|c|c|c|c|c|c|c|c|c|c|c|c|}
\hline
\multicolumn{4}{|c|}{-} & \multicolumn{5}{|c|}{two-flavored matter}  & \multicolumn{6}{|c|}{three-flavored  $\beta$-matter}  & same $\mu$      \\
\hline
B$^{1/4}$ (MeV)   & $X_V$  & $G_V$ (fm$^2$) & $b_4$   & $T_c$ (MeV) & $\mu_c$ (MeV) & Cleymans?& L-G? & $\mu_c$? & 
$M$ ($M_\odot$) & $R_{1.4}$ (km) & $T_{c(\beta)}$ & $\mu_{c(\beta)}$ & 1.96$M_\odot$? & $R_{1.4}$? & SQM?  \\
\hline
148 &  - & 0.0 & 0.0 & 121 & 938 & No & No  & No & 1.85 & 10.36 &   105   &  866 & No  &  No & Yes   \\
165 &  - & 0.0 & 0.0 & 134 & 1040& No & Yes & No & 1.50 &  8.58 &   117   &  962 & No  &  No & No  \\
205 &  -  & 0.0 & 0.0 & 168 & 1296& Yes& Yes& Yes& 0.98 & -     &   145  &  1187 & No  &  No & No   \\
\hline
148 & 1.0 & 0.3 & 0.0 & 121 & 1000& No & Yes & No & 2.36 & 11.15 &  105   &  938 & Yes &  Yes & No   \\
148 & 1.0 & 0.3 & 60  & 121 & 987 & No & Yes & No & 1.99 & 10.73 &  105   &  924 & Yes &  Yes & Yes   \\
165 & 1.0 & 0.8 & 0.0 & 134 & 1249& No & Yes & Yes& 2.31 & 10.20 &  117   &  1186 & Yes &  No  & No  \\
165 & 1.0 & 0.8 & 20  & 134 & 1163& No & Yes & Yes& 1.58 &  8.94 &  117   &  1091 & No  &  No  & No  \\
205 & 1.0 & 0.8 & 0.0 & 168 & 1662& Yes& Yes & No & 1.62 &  7.60 &  145   &  1575 & No  &  No  & No  \\
205 & 1.0 & 0.8 & 20  & 168 & 1464& Yes& Yes & No & 1.02 &  -    &  145   &  1364 & No  &   -  & No  \\
\hline
148 & 0.4 & 0.3 & 0.0 & 121 & 1000& No & Yes & No & 2.19 & 10.93 &  105   &  915 & Yes &  Yes  & Yes    \\
148 & 0.4 & 0.3 & 60  & 121 & 987 & No & Yes & No & 1.96 & 10.67 &  105   &  914 & Yes &  Yes  & Yes  \\
165 & 0.4 & 0.8 & 0.0 & 134 & 1249& No & Yes & Yes& 2.04 &  9.78 &  117   &  1111 & Yes &  No   &  No \\
165 & 0.4 & 0.8 & 20  & 134 & 1163& No & Yes & Yes& 1.56 &  8.86 &  117   &  1055 & No  &  No   &  No \\
205 & 0.4 & 0.8 & 0.0 & 168 & 1662& Yes& Yes & No & 1.41 &  6.72 &  145   &  1442 & No  &  No   &  No \\ 
205 & 0.4 & 0.8 & 20  & 168 & 1464& Yes& Yes & No & 1.00 &    -  &  145   &  1320 & No  &   -   &  No \\
\hline
\end{tabular}
}
\caption{Critical temperature ($T_c$), chemical potential ($\mu_c$), quark star main properties, and some observational constraints for two and three-flavored quark matter within different values of $B^{1/4}$, $G_V$, $b_4$ and $X_V$.} 
\label{T1}
\end{table}
\end{center}
\end{widetext}

From Fig.~\ref{F1} we can see that when there is no interaction term, increasing bag pressure values favor the hadron phase. The higher the $B$ value, the higher both the critical temperature and the critical chemical potential are. 
The vector field causes an additional repulsion, increasing the critical chemical potential. However, as shown in Eq.~(\ref{v3}), the vector channel couples to the density and causes very little effects at low chemical potential. Indeed, the vector field does not change the critical temperature for $\mu$ = 0. The self-interacting term reduces the repulsion 
at large densities. In our previous work~\cite{Carline}, we have used low values of $b_4$ to not change the stability window, but here we relax this condition in order to study the strong coupling effects.
We see that for $B^{1/4}$ = 148 MeV (the only value in this work with which it is possible to reproduce SQM) the critical temperature for zero chemical potential is 121 MeV. This value is 
way lower than the experimental results from chemical freeze-out~\cite{Cleymans} and the LQCD~\cite{Bazavov1,Bazavov2}. At the low temperature limit, we see that without the vector channel,
even the region related to the liquid-gas phase transition is not fully contained in the hadron phase. We overcome this issue by adding the vector field, but it is not enough to push
the critical chemical potential for values larger than 1000 MeV, which is still below the minimum of 1050 MeV.
For $B^{1/4}$ = 165 MeV, the critical temperature at zero chemical potential is 134 MeV, which is still not enough to match the LQCD and freeze-out results. 
Without the vector channel, the critical potential at zero temperature is 1040 MeV, but with $G_V$ as well as with the $b_4$ term, the critical chemical potential
is always in the range of $1050$ MeV $< \mu_c < 1400$ MeV.
Finally, for $B^{1/4}$ = 205 MeV we have both the freeze-out line and the LQCD results in agreement with our model, i.e. the lines are contained in the confined (hadronic) phase. But the vector channel pushes the
critical chemical potential for values above 1400 MeV.

 Although $B^{1/4}$ = 205 MeV seems to describe reasonably well the low chemical potential region of the QCD phase diagram, it can be seen from Fig.~\ref{F2} that it results in quark stars of very low masses. Indeed, in most cases
the maximum mass barely reaches 1.0$M_\odot$. With $B^{1/4}$ = 148 MeV we are able to reproduce a 2.19 $M_\odot$ quark star within the SQM conjecture and a 2.36 $M_\odot$ in
general. For $B^{1/4}$ = 165 MeV a maximum mass of 2.31 $M_\odot$ arises, a result very close to 2.36 $M_\odot$ with $B^{1/4}$ = 148 MeV. However, although the maximum masses are very close, there is a huge difference on the radii of the canonical stars. For $B^{1/4}$ = 148 MeV a radius of 11.15 km is obtained, while for $B^{1/4}$ = 165 MeV
the radius is only 10.20 km, about 1 km smaller. The vector channel can help to differentiate the values of the bag pressure, even with similar maximum masses. The radii are very sensitive to the bag pressure value. We also see that the maximum mass is sensitive to $X_V$ in absence of the self interacting term. Within a strong coupling constant $b_4$, the effect of $X_V$ is secondary. Indeed, the self-interacting term makes the results closer to the ones obtained with the non-interacting quark gas than the ones with the vector field only. Another important feature is the radius of the canonical star. We see that the only models whose radii
 are  10.4 km$<R_{1.4}<$11.9 km~\cite{Capano} are those
with bag equal to 148 MeV, which is within (or very close to) the
stability window of SQM~\cite{Carline}. Bag pressure values 
that are far from the stability
window  produce too low radii, even if their maximum masses are compatible with the MSP J070+6620~\cite{Cromartie}.

Finally, from Fig.~\ref{F3} we see that the $\beta$-stable matter disfavors the hadronic phase when compared 
with symmetric two-flavored matter at Fig.~\ref{F1}, as it always produces a lower value for the critical temperature (at zero chemical potential),
as well as a lower value for the critical chemical potential (at zero temperature). Also, when we compare $X_V$ = 1.0 to $X_V$ = 0.4
we see that higher values of $X_V$ push the critical chemical potential to higher values. 

The results of this section are
summarized in Tab.~\ref{T1}, where the critical temperature and chemical potential in two and three-flavored matter are presented,
the constraints discussed in the text from both freeze-out and neutron stars observations and what parametrizations give us a SQM for $\mu_u = \mu_d = \mu_s$.
We also need to notice that not all the EoS satisfy the Bodmer-Witten conjecture.

\section{Temperature-dependent bag model}

\begin{figure}[ht] 
\begin{centering}
 \includegraphics[angle=270,
width=0.49\textwidth]{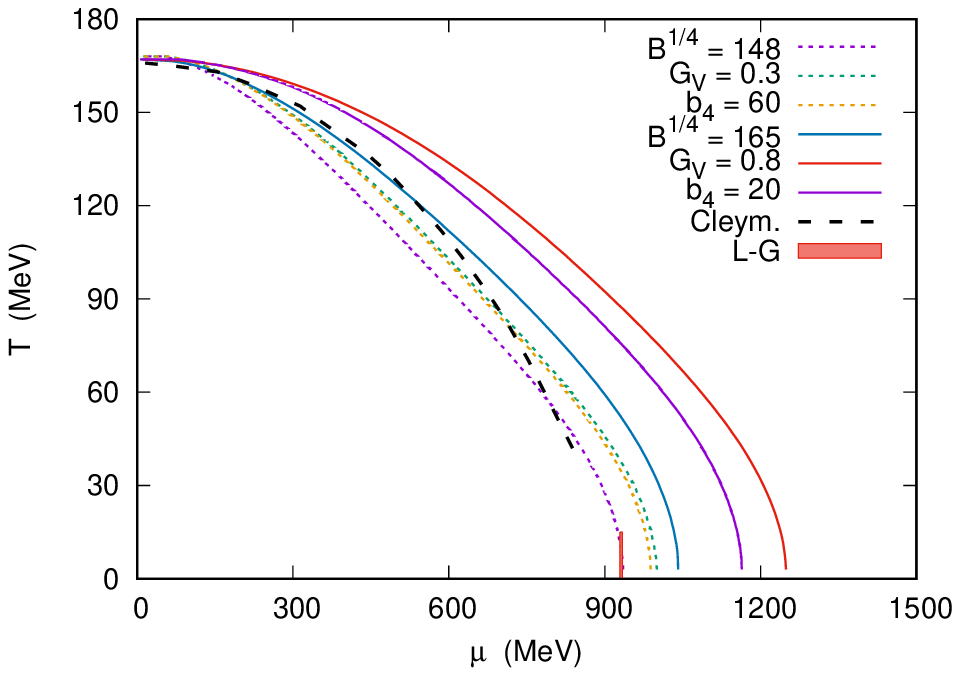}
\caption{(Color online) QCD phase diagram for temperature dependent bag model}\label{F4}
\end{centering}
\end{figure}

It is clear from the last section that we are not able to reproduce both the freeze-out  critical temperature and massive quark stars.
Even if we increase the value of  $G_V$ for $B^{1/4}$ = 205 MeV, it pushes the critical chemical potential even further, in disagreement with the probable existence of a
quark core as presented in ref.~\cite{Nature2020}, and reproduces quark stars with too low radii, in disagreement with
the measurements from ref.~\cite{Capano}.  Because of these behaviors, different bag pressure values are generally used for different energy scales. When the interest lies in the low chemical potential, a bag pressure value higher than 200 MeV is employed, while for constructing quark stars, values around 150-170 MeV are generally used~\cite{Rafa2011,Pedro}. Hence, it 
makes it impossible to use a single bag pressure value to describe the whole energy spectrum with MIT-based models.
We then propose an alternative, a temperature dependent bag model, a prescription that is not new. For instance, ref.~\cite{Bhalerao} uses a temperature dependent bag model derived from QCD sum rules and ref.~\cite{Muller,Dey} introduce a phenomenological temperature dependence in the bag constant,
which reads:

\begin{equation}
B =  B_0 \bigg [1 - \bigg (\frac{T}{T_0} \bigg )^4 \bigg ] , \label{deybag}
\end{equation}
where $T_0$ is a free parameter. 
However, as noted in ref.~\cite{Prasad}, this bag model is the thermodynamic potential itself: $B = \Omega$. Another possibility is a density-dependent bag model, as proposed in ref.~\cite{mallick2013}. Notice that in  ref.~\cite{Song}, the authors introduce a bag parametrization that depends on both, temperature and chemical potential.

Here we propose a bag model similar to ref.~\cite{Dey}, but where the bag increases with the temperature instead of decreasing: 
\begin{equation}
B(T) =  B_0\bigg [1 + \bigg (\frac{T}{T_0} \bigg )^4 \bigg ] . \label{ourbag}
\end{equation}

\begin{figure*}[htb]
\begin{tabular}{cc}
\includegraphics[angle=270,width=0.49\textwidth]{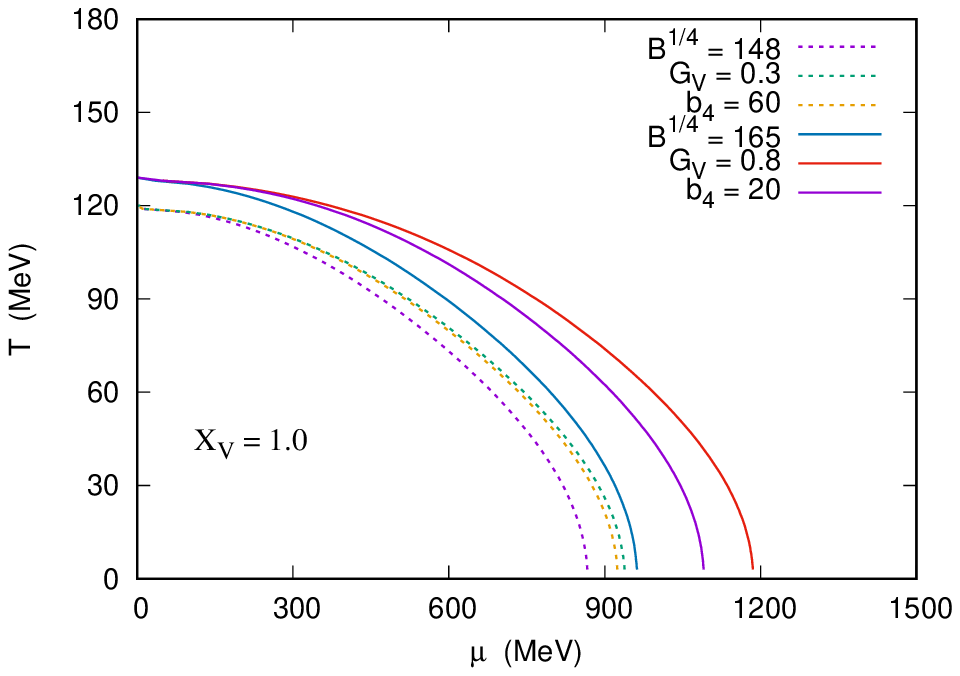} &
\includegraphics[angle=270,width=0.49\textwidth]{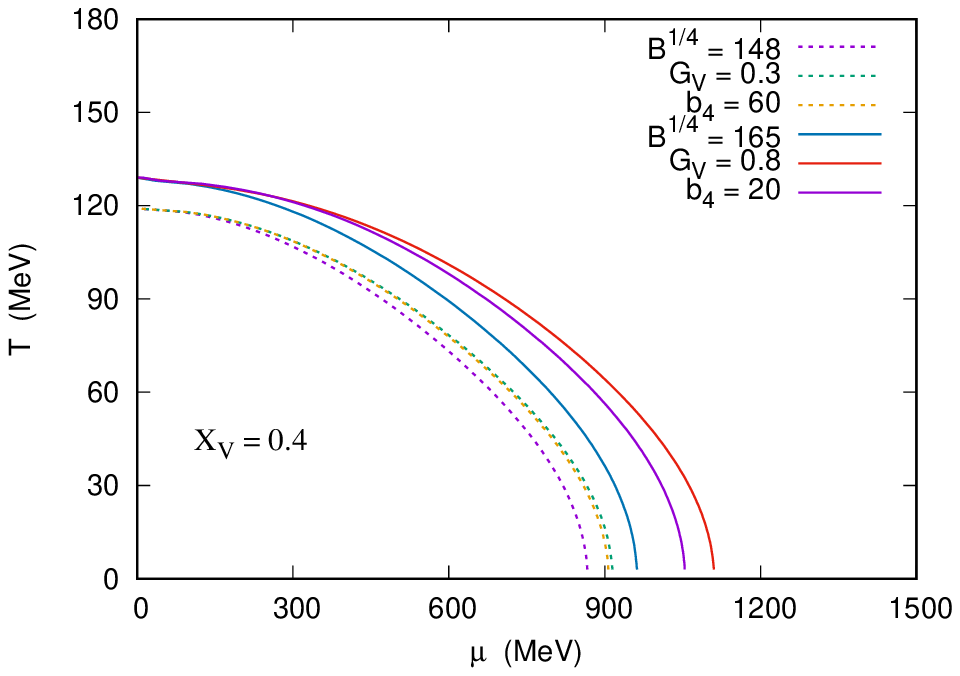} \\
\end{tabular}
\caption{(Color online) Phase diagram for $X_V$ = 1.0 (left) and $X_V$ = 0.4 (right) for neutral beta stable matter within B(T) formalism. } \label{F5}
\end{figure*}

Now $T_0$ is adjusted to reproduce the
LQCD and freeze-out (pseudo) critical temperature at zero chemical potential.
Therefore $T_0$ = 131 MeV for $B_0^{1/4}$ = 148 MeV, and $T_0$ = 155 MeV
for $B_0^{1/4} = 165$ MeV. The increase of the bag pressure value at high temperature
can be justified not only to reproduce the chemical
freeze-out results~\cite{Cleymans}, but also to increase the quark surface at high temperatures. As shown
in ref.~\cite{surface}, the surface tension above 100 MeV  increases monotonically with $T$.
This model decouples the bag parameter from the thermodynamic potential, and two distinct couplings appear: the vector field only couples to the density while the bag parameter only couples to the temperature.

The QCD phase diagram with two-flavored $\mu_u =\mu_d$ matter is 
presented in Fig.~\ref{F4}. As can be seen, we obtain the critical temperature
$T_c$ around  168 MeV for both $B_0$ values. However, for $B^{1/4}_0$ = 148 MeV, the freeze-out line is not entirely inside the confined (hadron) phase. Even for the upper limit of the stability window we can't satisfy this restrain. This
indicates that, for this model, we are not able to reproduce both,
SQM and the QCD phase diagram. Also, as for temperatures $T \simeq 0$, so that $B(T) \simeq B_0$,
this bag pressure value produces a low critical chemical potential. 
On the other hand, when we use $B^{1/4}_0$ = 165 MeV, all the freeze-out
region lies within the hadronic phase, except for the non-interacting case. 
Also, the critical chemical potential at zero temperature lies between 1050 MeV $<\mu_c<$ 1400 MeV,  resulting on
a good description of the QCD diagram. In the
same sense, as shown in Fig.~\ref{F2}, we are able to reproduce
massive stars at $T = 0$ approximation, in agreement of ref.~\cite{Cromartie}. Nevertheless,
this bag pressure value is far above the stability window, even for non-interacting
bag model.

In Fig.~\ref{F5} we display the phase diagram for tree-flavored 
$\beta$-stable matter for $X_V$ = 1.0 and $X_V$ = 0.4. For low temperatures
Fig.~\ref{F5} and Fig.~\ref{F2} are similar. But at low chemical potential
we have a increase of the critical temperature due to the dependence of the bag on the
temperature. When compared with two-flavored symmetric matter,
there is a decrease from the critical temperature at low chemical potential,
as expected. All the numerical values are presented in Tab.~\ref{T2}.
As can be seen, for $B_0^{1/4} = 165$ MeV, with the linear vector channel, both values of $X_V$
are able to simultaneously give a good description of the QCD phase diagram and to reproduce
massive quark stars. However these results point against the existence of SQM, as they are 
far from the stability window. This is reinforced by the low radii of the canonical stars,
which are in disagreement with ref.~\cite{Capano}, indicating that canonical stars are probably
hadronic neutron stars.

\section{Hot quark stars and the speed of sound}

\begin{figure*}[htb]
\begin{tabular}{cc}
\includegraphics[angle=270,width=0.49\textwidth]{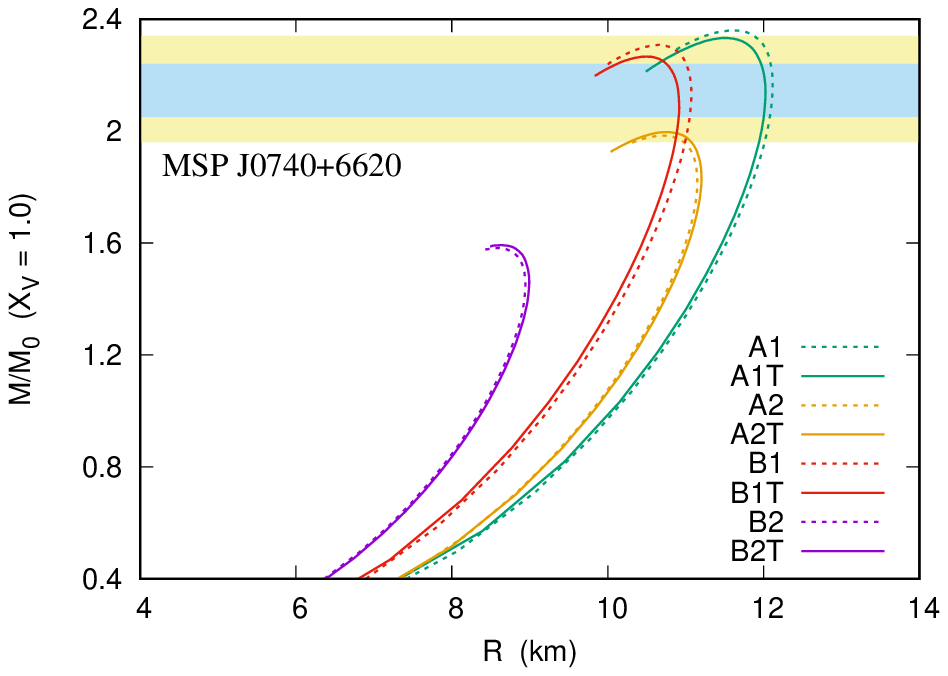} &
\includegraphics[angle=270,width=0.49\textwidth]{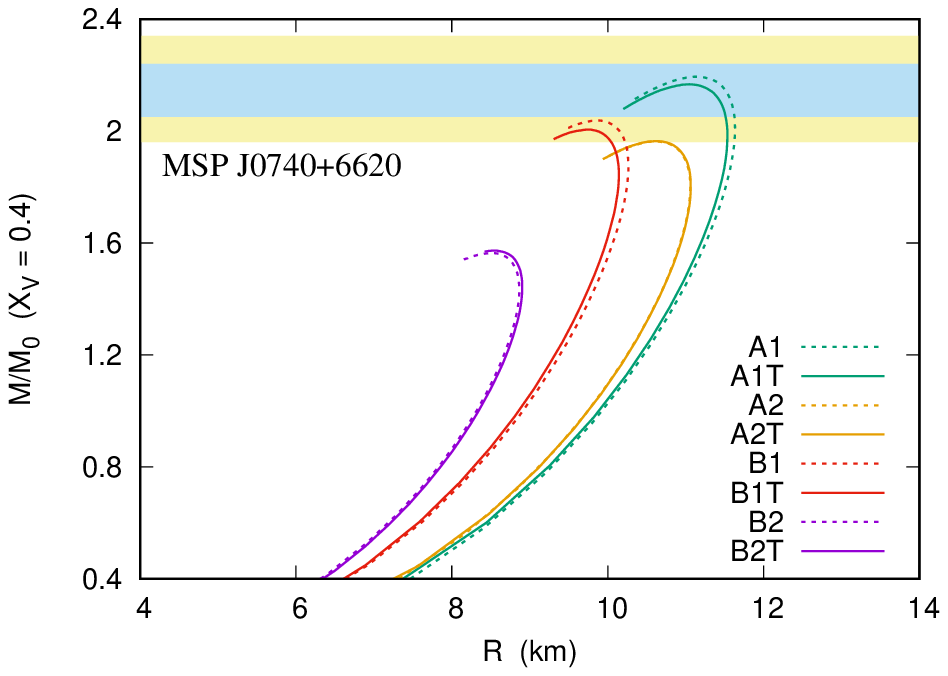} \\
\end{tabular}
\caption{(Color online) Mass-radius relation for T = 0 and T = 40 MeV, with different approaches.
We have used the following notation for labelling the curves: A means $B^{1/4}$ = 148 MeV, B means
$B^{1/4}$ = 165 MeV. The number 1 means we include a linear coupling with $G_V$,
and the number 2 implies the existence of the linear and the self-interaction
$b_4$.  The results for $T = 40$ MeV
are represented by the letter T. T = 0 approximation has no subscript. } \label{F6}
\end{figure*}

As pointed out in ref.~\cite{Lattimer,Lattimer2}, a newborn neutron star is formed 
in the aftermath of a successful supernova explosion as the stellar remnant
becomes gravitationally decoupled from the expanding ejecta, reaching 
a temperature as high as 500 billion Kelvin in the core (around 50 MeV).
And while it takes a dozen of seconds to cool down, deconfinement and
chemical equilibrium are reached in a timescale millions and millions of times
lower~\cite{Benvenuto}. Therefore the study of hot quark matter
is important in both macroscopic properties of quark stars as
well as to how the temperature affects the critical chemical potential.

Here we study how a temperature of 40 MeV affects the mass-radius
relation of quark stars. Although some studies point out that an isentropic
formulation is more realistic~\cite{Lattimer2,Mal,Debora2004}, we choose
an isothermic one, as made in previous works~\cite{Gupta,Rafa2008,Santos2004,Jia,Bordbar}, which
allows a direct comparison with the phase diagram  presented in Fig.~\ref{F2}.
We also use $B_0$ instead of B(T), as for 40 MeV the increase of the bag pressure value
is always below 1$\%$ (even below 0.5$\%$ for $B^{1/4}$ = 165 MeV.)
In Fig.~\ref{F6} the mass-radius relation is presented 
 for $B^{1/4}$ = 148 MeV and $B^{1/4}$ = 165 MeV with $X_V$ = 1.0
and $X_V$ = 0.4.

\begin{figure*}[htb]
\begin{tabular}{cc}
\includegraphics[angle=270,width=0.49\textwidth]{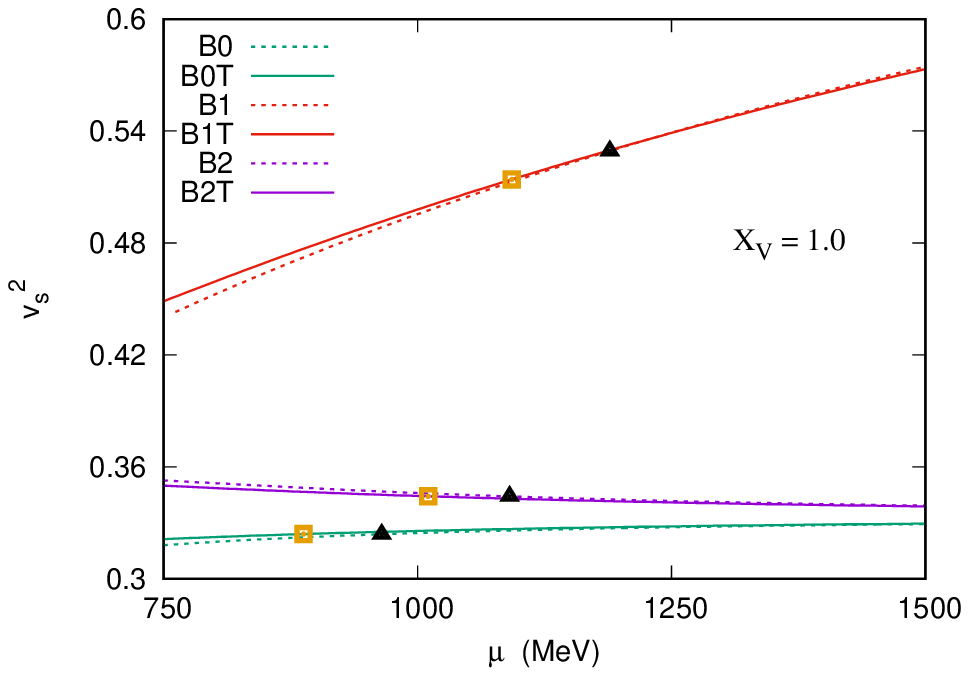} &
\includegraphics[angle=270,width=0.49\textwidth]{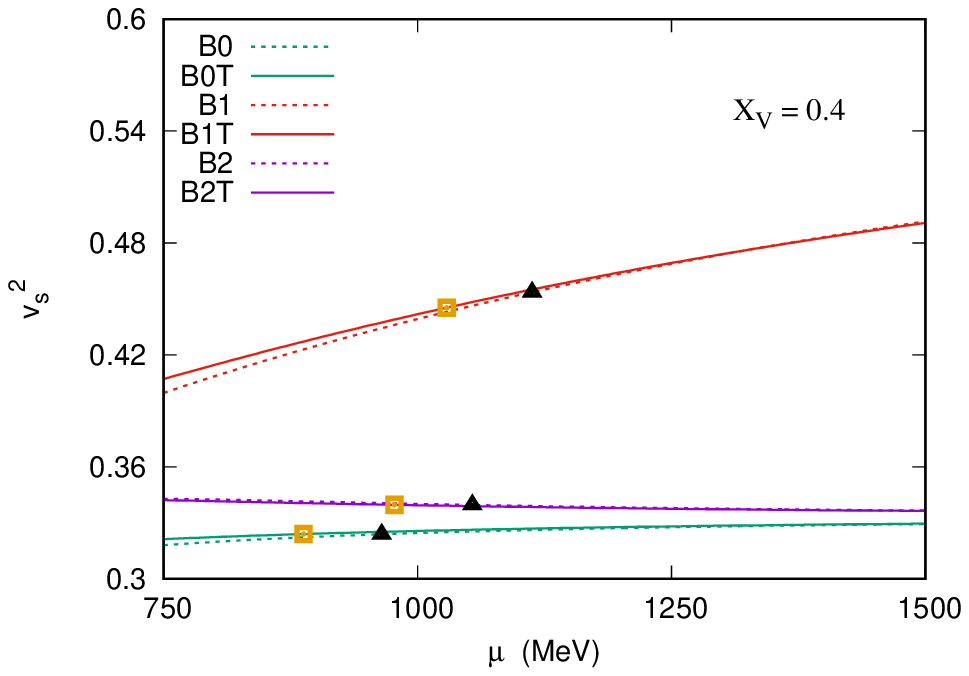} \\
\end{tabular}
\caption{(Color online) Speed of the sound of quark matter for $B^{1/4} = 165$ MeV with $X_V$ = 1.0 (left) and $X_V$ = 0.4 (right). The notation for the labels is the same as the one used in Fig. \ref{F6} and 0 means non interacting bag model. 
The black triangles represent the critical chemical potentials for T = 0, while yellow squares represent the critical chemical potential for T = 40 MeV}.\label{F7}
\end{figure*}

\begin{widetext}
\begin{center}
\begin{table}[ht]
\centering
\resizebox{\textwidth}{!}{%
\begin{tabular}{|c|c|c|c|c|c|c|c|c|c|c|c|c|c|c|c|}
\hline
\multicolumn{4}{|c|}{-} & \multicolumn{5}{|c|}{two-flavored matter}  & \multicolumn{6}{|c|}{three-flavored  $\beta$-matter}  & same $\mu$      \\
\hline
B$^{1/4}$ (MeV)   & $X_V$  & $G_V$ (fm$^2$) & $b_4$   & $T_c$ (MeV) & $\mu_c$ (MeV) & Cleymans?& L-G? & $\mu_c$? & 
$M$ ($M_\odot$) & $R_{1.4}$ (km) & $T_{c(\beta)}$ & $\mu_{c(\beta)}$ & 1.96$M_\odot$? & $R_{1.4}$? & SQM?  \\
\hline
148 &  - & 0.0 & 0.0 & 168 & 938 & No & No  & No & 1.85 & 10.36 &   118   &  866 & No  &  No & Yes   \\
165 &  - & 0.0 & 0.0 & 167 & 1040& No & Yes & No & 1.50 &  8.58 &   128   &  962 & No  &  No & No  \\
\hline
148 & 1.0 & 0.3 & 0.0 & 167 & 1000& No & Yes & No & 2.36 & 11.15 &  118   &  938 & Yes &  Yes & No   \\
148 & 1.0 & 0.3 & 60  & 167 & 987 & No & Yes & No & 1.99 & 10.73 &  128   &  924 & Yes &  Yes & Yes   \\
165 & 1.0 & 0.8 & 0.0 & 168 & 1249& Yes & Yes & Yes& 2.31 & 10.20 &  118  &  1186 & Yes &  No  & No  \\
165 & 1.0 & 0.8 & 20  & 168 & 1163& Yes & Yes & Yes& 1.58 &  8.94 &  128  &  1091 & No  &  No  & No  \\
\hline
148 & 0.4 & 0.3 & 0.0 & 167 & 1000& No & Yes & No & 2.19 & 10.93 &  118   &  915 & Yes &  Yes  & Yes    \\
148 & 0.4 & 0.3 & 60  & 167 & 987 & No & Yes & No & 1.96 & 10.67 &  118   &  914 & Yes &  Yes  & Yes  \\
165 & 0.4 & 0.8 & 0.0 & 168 & 1249& Yes & Yes & Yes& 2.04 &  9.78 &  128  &  1111 & Yes &  No   &  No \\
165 & 0.4 & 0.8 & 20  & 168 & 1163& Yes & Yes & Yes& 1.56 &  8.86 &  128  &  1055 & No  &  No   &  No \\
\hline
\end{tabular}
}
\caption{Critical temperature ($T_c$), chemical potential ($\mu_c$), quark star main properties, and some observational constraints for two and three-flavored quark matter within different values of $B^{1/4}$, $G_V$, $b_4$ and $X_V$ within B(T) formalism.} 
\label{T2}
\end{table}
\end{center}
\end{widetext}

From Fig.~\ref{F6} we see that $B^{1/4}$ = 148 MeV and $B^{1/4}$ = 165 MeV,
have the same qualitative behavior for all $G_V$, $b_4$ and $X_V$.
In a non-interacting gas (not shown in order not to saturate the figure) we have just a small increase of the maximum mass for T = 40 MeV when compared to T = 0. For a linear  coupling, there is a more significant decrease of the mass instead of an increase. But when we include the vector with self-interaction we recover an increase of the mass with the temperature.
The decrease of the maximum mass within a linear vector field is due to the reduction of the central density with the temperature. 
The main contribution of the temperature is the reduction of the critical chemical potential.
In all cases studied, the critical chemical potential has a reduction around 7 - 10$\%$.

In Fig.~\ref{F7} we show the square of the speed of sound, $v_s^2$, of quark matter as a function of the chemical potential for $B^{1/4}$ = 165 MeV
(as the speed of sound does not depend on the bag pressure value, the qualitative results are maintained),
\begin{equation}
 v^2_s = \frac{\partial p}{\partial \epsilon}. \label{vsom}
\end{equation}

As recently pointed out in ref.~\cite{Nature2020} the speed of sound of the quark matter is closely related to the 
mass and radius of the quark core in hybrid stars. The authors found that if the
conformal bound ($v^2_s < 1/3)$ is not strongly violated, massive neutron stars are predicted to have sizable quark-matter cores. We plot the black triangles  (for T = 0) and yellow squares (for T = 40 MeV) that represent the critical chemical potential of each model. 
For the linear vector field, we obtain a monotonically increase of the speed of sound with the chemical potential. Also, the model with linear vector field
presents both the higher speed of sound and higher critical chemical potential. For the non-interacting
bag model, as well as for the bag model with self-interacting vector field we have an almost constant
speed of the sound. However the vector field displaces the speed of the sound to slightly higher values.
The non-interacting bag model produces the lower speed of the sound and the
lower critical potential. When we compare $X_V$ = 1.0 to $X_V$ = 0.4 we see that
only with the linear vector field there is a significant change in the speed of the sound. The self-interacting
field washes out the role of $X_V$. 
The main results of this section are presented in Tab.~\ref{T3}.

\begin{widetext}
\begin{center}
\begin{table}[ht]
\centering
\begin{tabular}{|c|c|c|c|c||c|c|c|c|c|}
\hline
B$^{1/4}$ (MeV)   & $X_V$  & $G_V$ (fm$^2$) & $b_4$   & $T$ (MeV) & $\mu_c$  (MeV)  &M  ($M_\odot$) & $v_s^2$ at $\mu_c$  \\
\hline
148 &  - & 0.0 & 0.0 & 0 & 866 & 1.85 & 0.32    \\
148 &  - & 0.0 & 0.0 & 40 & 781 & 1.86 & 0.32    \\
165 &  - & 0.0 & 0.0 & 0 &  962 & 1.50 & 0.32  \\
165 &  - & 0.0 & 0.0 & 40 & 885 & 1.52 & 0.32    \\
\hline
148 & 1.0 & 0.3 & 0.0 & 0 & 938 & 2.36 & 0.42   \\
148 & 1.0 & 0.3 & 0.0  & 40 & 847 & 2.33 & 0.40    \\
148 & 1.0 & 0.3 & 60 & 0 & 924 & 1.99 & 0.35   \\
148 & 1.0 & 0.3 & 60 & 40 & 836 & 2.00 & 0.35   \\
165 & 1.0 & 0.8 & 0.0 & 0 & 1186& 2.31 & 0.53  \\
165 & 1.0 & 0.8 & 0.0 & 40 & 1096 & 2.26 & 0.51  \\
165 & 1.0 & 0.8 & 20  & 0 & 1091& 1.58 & 0.34   \\
165 & 1.0 & 0.8 & 20 & 40 & 1009 & 1.59 & 0.34  \\
\hline
148 & 0.4 & 0.3 & 0.0 & 0 & 915& 2.19 & 0.39     \\
148 & 0.4 & 0.3 & 0.0 & 40 & 826 & 2.16 & 0.37  \\
148 & 0.4 & 0.3 & 60  & 0 & 914 & 1.96 & 0.34     \\
148 & 0.4 & 0.3 & 60  & 40 & 819 & 1.98 & 0.34  \\
165 & 0.4 & 0.8 & 0.0 & 0 & 1111& 2.04 & 0.46  \\
165 & 0.4 & 0.8 & 0.0 & 40 & 1249& 2.00 & 0.45  \\
165 & 0.4 & 0.8 & 20  & 0 & 1055 & 1.56 & 0.34  \\
165 & 0.4 & 0.8 & 20  & 40 & 974 & 1.56 & 0.34 \\
\hline
\end{tabular}
\caption{Critical chemical potential ($\mu_c$), maximum mass and speed of the sound in function of several parameters, as discussed in the text.} 
\label{T3}
\end{table}
\end{center}
\end{widetext}

\section{Comparison with the Nambu--Jona-Lasinio model}

 In this section we also reproduce some results obtained with the 
Nambu--Jona-Lasinio (NJL) model to compare with the ones we have obtained in the present work.
Although the simplest versions of the NJL model does not reproduce the asymptotic freedom behavior of QCD, and thus cannot describe the quark confinement/deconfinement transition, it can be interpreted as a schematic quark model for many situations where chiral symmetry breaking/restoration is one of the most relevant features of QCD \cite{Buballa2005}. 

The original version of the model \cite{Nambu} can be extended by the inclusion of a vector-isoscalar interaction term, along the same lines discussed in ref.~\cite{Carline} and written in Eq.(\ref{v0}) for the MIT bag model,
such that, considering two quark flavor fields $\psi_q=\left [\psi _u\; \; \; \psi_d  \right ]^T$, it is  given by the following Lagrangian density
\begin{align}
        \mathcal{L} ={}& \bar{\psi}_q(i\gamma^\mu\partial_\mu - \hat m)\psi_q+ G_s[(\bar{\psi}_q\psi_q)^2 + (\bar{\psi}_qi\gamma_5\vec{\tau}\psi_q)^2] \nonumber\\&- G_v(\bar{\psi}_q\gamma^\mu \psi_q)^2,\label{eq:njlsu2}
\end{align}
where $\hat m={\rm diag}(m_u,m_d)$ are the quark bare masses, $\vec{\tau}$ is the Pauli isospin matrix, and $G_s$ and $G_v$ are the coupling constants.
In the mean-field level, Lagrangian (\ref{eq:njlsu2}) can be rewritten via the bosonization of the model through auxiliary fields given by the non-vanishing scalar and vector condensates, from where it is straightforward to write the grand-canonical thermodynamic potential as~\cite{Buballa2005}:
\begin{equation} 
\Omega =\Omega _M+\frac{\left ( M-m \right )^2}{4G_s}-\frac{\left ( \mu_q -\tilde{\mu}_q \right )^2}{4G_v}, \label{eq:omega}
\end{equation}
with
\begin{align}
\Omega _M=-2N_c\int \frac{d^3k}{(2\pi)^3}\Big\{ E_k
&{}+T\ln\left [ 1+e^{ -(E_k-\tilde{\mu}_q )/T} \right ] \nonumber\\
&{}+\left. T\ln\left [ 1+e^{ -(E_k+\tilde{\mu}_q )/T} \right ] \right  \}, 
 \end{align}
being the displaced Fermi gas contribution, where $E_k = \sqrt{k^2 + M^2}$. In this process, the constituent mass and effective chemical potential were introduced, respectively, through the self-consistent gap equations~\cite{Buballa2005}:
\begin{align} 
M={}&m-2G_s\langle \bar{\psi}_q\psi_q\rangle,
\label{eq:gaprho}\\
\tilde{\mu}_q={}&\mu_q -2G_v\langle\bar{\psi_q}\gamma^0\psi_q\rangle. \label{eq:gaprho2}
\end{align}
The vector condensate is identified as the number density $\langle\bar{\psi_q}\gamma^0\psi_q\rangle=n_q$, as in equation (\ref{nd}), and the scalar condensate can be evaluated applying standard techniques of thermal field theory from the dressed fermion propagator $S= (\gamma^\mu k_\mu-M+i\varepsilon)^{-1}$ as
\begin{equation}
\left \langle \bar{\psi}_q\psi_q  \right \rangle=-2\int \frac{d^3k}{\left ( 2\pi  \right )^3}\frac{M}{\sqrt{k^2+M^2}}\left ( 1-f_{q+} - f_{q-}\right ).
\end{equation}
Here the isospin symmetry is assumed in the Lagrangian level, i.e., $m_u=m_d=m$, and the situation of chemical equilibrium is considered, i.e., $\mu_u = \mu_d = \mu_q$, with $\mu_q= \mu/3$. The parameters $G_s$,  $m$ and $\Lambda$ are fitted to reproduce the quark condensate value, the pion mass and its decay constant. In the following, we set $\Lambda = 590$ MeV, $G_s\Lambda^2 = 2.435$ and $m = 6.0$ MeV \cite{Buballa2005}. The parameter $G_v$ is usually taken as a free parameter, whith some proposed constraints suggesting values between $0.25 G_s$ and $0.5 G_s$ \cite{gv1,gv2}. We  consider $0\leq G_v/G_s\leq 0.5$.

\begin{figure}[ht] 
\begin{centering}
 \includegraphics[angle=270,
width=0.49\textwidth]{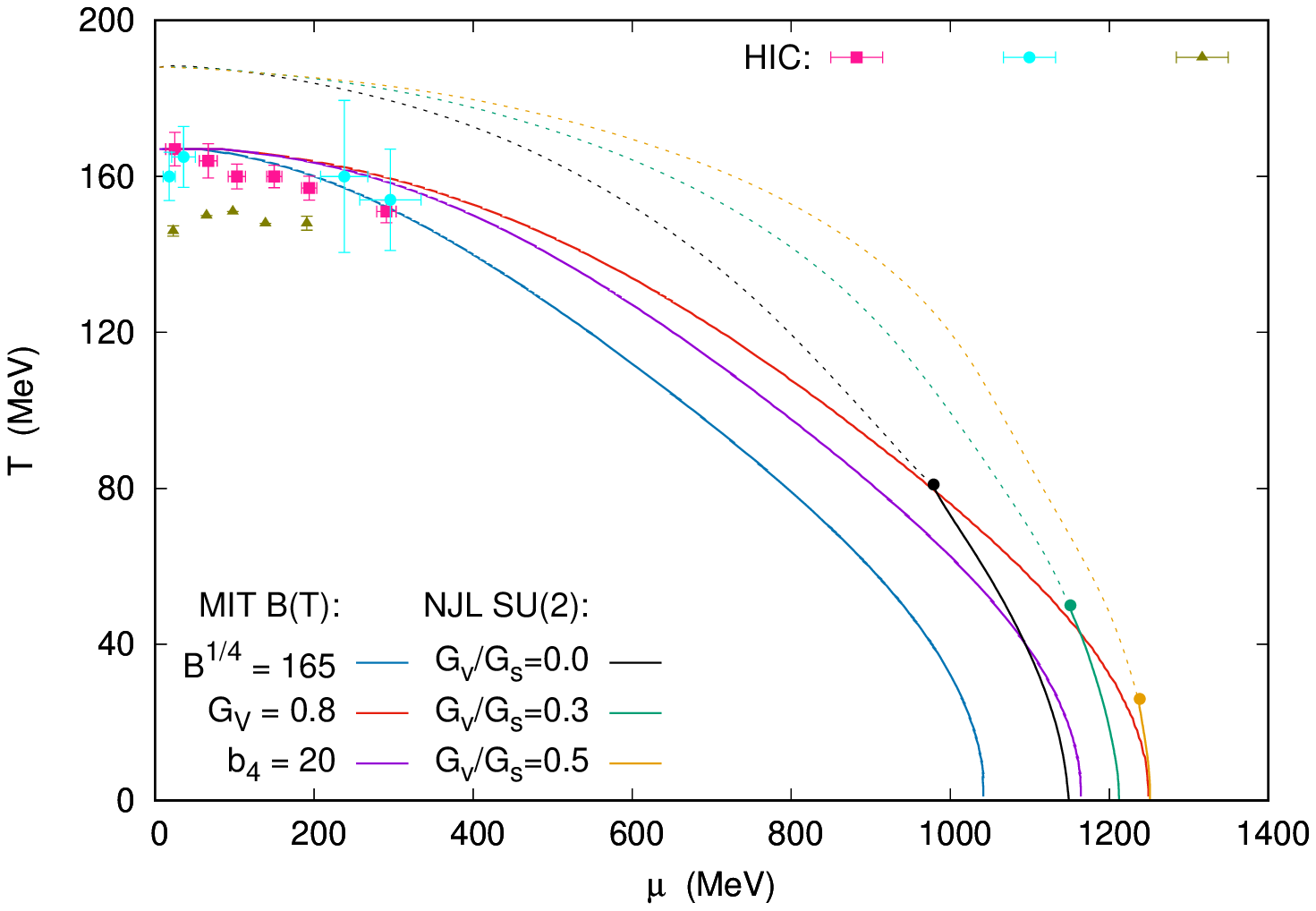}
\caption{(Color online) QCD phase diagram for different values of $G_v/G_s$, from the SU(2) NJL model \cite{Robson}. A solid
line represents a first order transition and a dashed line represents a crossover, the intersection is the CEP. Selected curves from the temperature dependent bag model are included for comparison, as well as some estimations of the chemical freeze-out parameters in heavy ion collisions \cite{hicpink,hicblue,hicbrown}. }\label{Fk}
\end{centering}
\end{figure}

\begin{figure}[ht] 
\begin{centering}
 \includegraphics[angle=270,
width=0.49\textwidth]{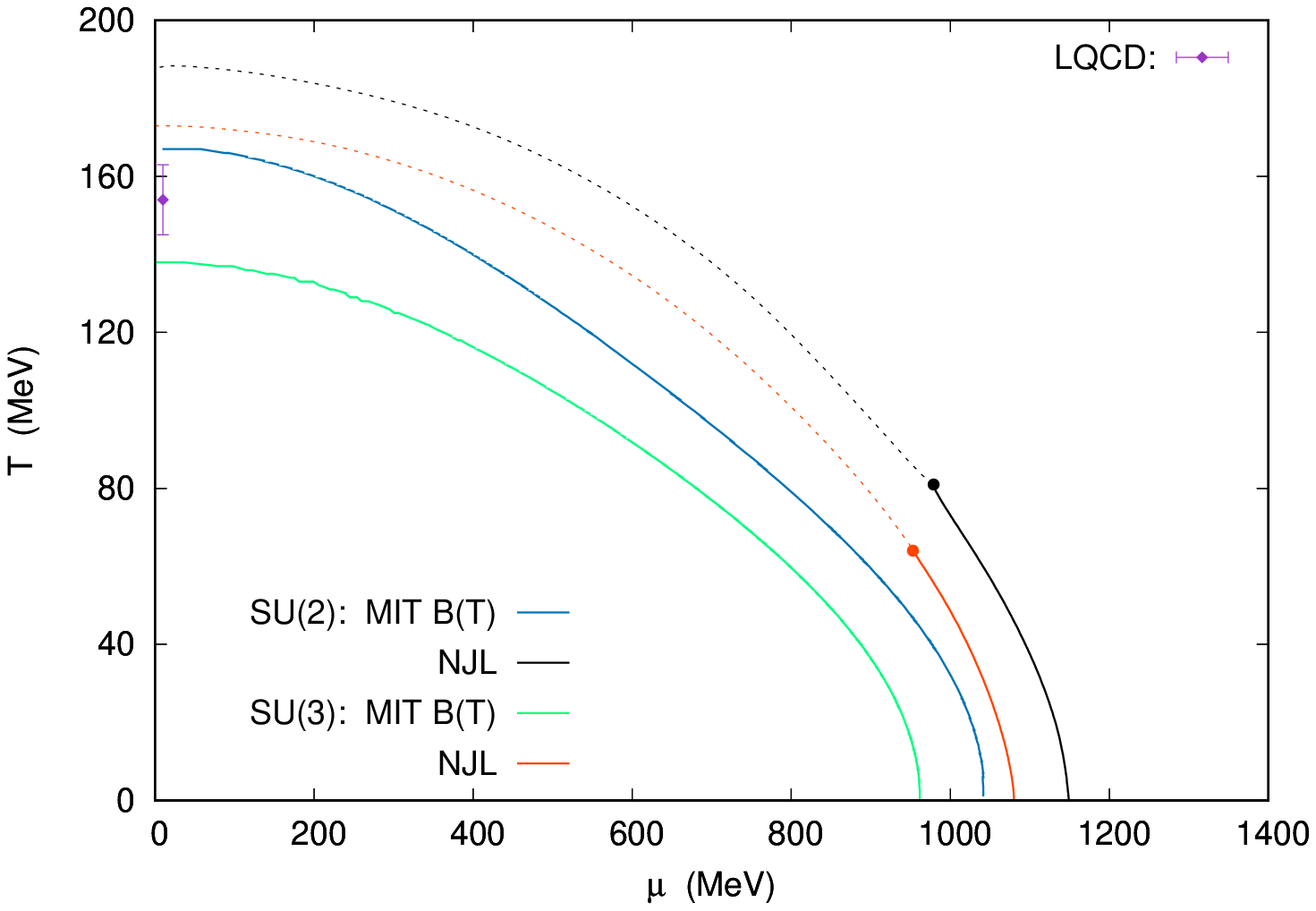}
\caption{(Color online) QCD phase diagram for SU(2) and SU(3) NJL model ($G_v/G_s=0.0)$ \cite{Robson,PRD85} and temperature dependent bag model ($B^{1/4}$ = 165 MeV and $G_V=0$). A solid
line represents a first order transition and a dashed line represents a crossover, the intersection is the CEP. The LQCD predicted critical temperature for 2+1 quark flavors \cite{lqcdtemp} is included for comparison. }\label{Fk2}
\end{centering}
\end{figure}

The effective quark mass $M$ is obtained solving the gap equations (\ref{eq:gaprho}-\ref{eq:gaprho2}). This constituent mass is larger than the bare quark mass $m$ at lower temperatures and/or densities, generating dynamically the larger particle mass expected in this region and breaking the chiral symmetry of the model. As the temperature or the density increase, $M$ approaches the value of the current mass $m$, thus restorating the chiral symmetry. A chiral phase transiton $\mu-T$ diagram can be drawn determining the behavior of the thermodynamic potential minima with respect to $M$, for given chemical potentials \cite{Robson}. In the low temperature regime, several effective models predict a first order chiral phase transition to occur. Results from LQCD for the low chemical potential region, however, point to a crossover transition. These two seemingly contradictory pictures suggest that the first order transition line starting at $T = 0$ ends at a critical end point (CEP), from which it turns into a crossover \cite{NJLv_CEP}. Figures \ref{Fk} and \ref{Fk2} show that this behavior is reproduced by the NJL model. Notice that, as the contribution associated to the vectorial coupling vanishes at zero chemical potential, there is no vacuum correction consequences due to the value of the coupling constant $G_v$. With the chosen parametrization, this model renders the crossover temperature at $\mu=0$ being equal to 188 MeV, which is higher than the values obtained from previously discussed MIT-type model calculations, but also from the estimations of the chemical freeze-out parameters in heavy ion collisions and expected from LQCD results, as can be seen in Figs.~\ref{Fk} and \ref{Fk2}.
As stated above, the NJL model produces a first order phase transition for temperatures below the CEP, where it acquires a second order phase transition point, before the crossover region.  Also, increasing the vector term weakens and delays the first order phase transition of the chiral restoration, favoring the crossover transition on the majority of the QCD phase diagram high temperature-low baryonic density part. 

The purpose of the present section is to compare different model frameworks, and hence it is useful to display both two and three-flavored matter results obtained from the NJL-type models too. The extension of the NJL SU(2) to NJL SU(3) is not as straightforward as the inclusion of the $s$ quark in the MIT-like models. Thus, we refer the interested reader to refs.  \cite{Buballa2005}. The expressions for the grand-canonical potentials, the related gap equations and the chosen parametrizations are the same as in ref. \cite{PRD85}, taking $\Lambda = 631.4$ MeV, $G_s\Lambda^2 = 1.835$, $K\Lambda^5 = 9.29$, $m_u=m_d = 5.5$ MeV and $m_s135.7$ Mev.
 As extensively discussed in \cite{Muller}, the strange quark is largely responsible for
shaping the phase diagram of QCD
since its mass controls the nature of the chiral and deconfinement transitions. In ref.~\cite{Fukushima}, the schematic figure of the Columbia phase diagram in 3-flavour QCD points to the importance of the strange quark mass, which also plays a decisive role on the possible existence of a quark core in collapsed stars.
Hence, to consider strangeness conservation without relying on the inclusion of $\beta$-stable matter, which, as discussed below, is justified only at the lower temperature range of the QCD phase transition diagram, we show the curves obtained with the same 
chemical potential for the three quarks, i.e., $\mu_u=\mu_d=\mu_s$
in Fig. \ref{Fk2}, for the NJL and the MIT models. One can clearly see that within the SU(3) version of both models, the transition temperature at low chemical potential is lowered considerably in relation to the SU(2) curve and at low temperatures, the transition chemical potentials are also shifted towards lower values.
Also, the effects of the vector interaction are the same as the ones just discussed for the SU(2) case, e.g. the transition point at $\mu=0$ would be also kept fixed had different values of the vector interaction been plotted, as explained above.

Analyzing this framework and the results presented in previous sections together, it is possible to argue that the phase transition in QCD can take place either in one or two different steps, depending on the parameter choice adopted for the MIT-type model. From Figures \ref{F1} and \ref{F4}, we can see that both MIT models allow the deconfinement phase transition to take place around $\mu=1200$ MeV in the low temperature region, at least for some sets. If this is the case considered, it suggests that both deconfinement and chiral transition occur simultaneously in the QCD phase diagram. It does happen at $T=0$, e.g., when the parameters of the MIT model are taken to be $B^{1/4}$ = 165 MeV and $G_V=0.8$ fm$^2$, while the vector coupling of the NJL is set as $G_v/G_s=0.5$, at $\mu=1250$ MeV. However, even in such case, the MIT and NJL transition curves diverge rapidly for finite temperatures, as the dependence of the transition temperature on the chemical potential is noticeably more intense in the MIT curves, specially for chemical potential greater than the $\mu$ at the CEP predicted for the NJL-type models. 

The description of stellar matter cannot be done using the two-flavor formalism presented in this section, since strangeness is necessary to fulfill the Bodmer-Witten conjecture (but it is still a possibility  in the framework of quark-meson models, as pointed out in ref.~\cite{Chen}).
Both hybrid and quark stars with vector NJL models have been described in several studies \cite{Lopes2020,Bla,Lenzi,Shao,Tk,Ciern,Sandoval}. It is worth noting, however, that these models do not produce stable quark matter at zero temperature and/or magnetic field \cite{njlstab},  but they can certainly describe the inner matter of a hybrid star
\cite{Nature2020}, which is enough to justify the application of this type of model in theoretical studies, mainly the ones involving phase transitions. Increasing the vector term stiffens the equation of state, thus sustaining larger maximum stellar masses, but the macroscopic properties of the compact star depend strongly on the remaining parameter choice. As example, models used in ref. \cite{NJLv_stars} do not reach the 1.96 $M_\odot$ even for higher vector couplings ($G_v/G_s> 0.6$).

\section{Final remarks}

  In this work we have extended the modified versions of the MIT bag model, recently introduced in ref.~\cite{Carline}, on which a vector field and a self-interacting term are added, to consider finite temperature effects and then build the QCD phase diagram.
We first consider
symmetric two-flavored quark matter constrained to both the freeze-out and the
liquid-gas phase transition at the hadronic phase. Then, we have verified that it is difficult to reconcile these constraints with the existence of stable strange quark matter as proposed by the Bodmer-Witten
conjecture~\cite{Bod,Witten}.

We have next constructed the QCD phase diagram for three-flavored quark matter in $\beta$ equilibrium and zero electrical charge with
two different possibilities
for the strength of the strange quark interaction with the vector field: an
universal coupling, when the strength of the $s$ quark is equal to the $u$ and $d$ quarks, and one coming from the group theory approach
that fixes the $s$ quark coupling constant to 2/5 of the $u$ and $d$ quarks~\cite{Carline}. These conditions were then utilized to obtain EoS at zero and finite temperature and used as input for the TOV equations and to compute the sound velocity. The related mass-radius diagrams were displayed and discussed. Within different choices of parameters, massive stars can be described, but the radii of the canonical stars are generally too small, indicating that they are more likely to be hadronic stars.

Finally, we have revisited QCD phase diagrams obtained with the Nambu-Jona-Lasinio model with and without a vector interaction. Despite the fact that the nature of the transition described by the MIT-like and NJL models framework is different, we have seen that both formalisms can produce transition curves with the same overall behavior as the inclusion of the vector interaction is concerned, specially in the low temperature region.
Moreover, depending on the parameters choice for both models, the NJL chiral symmetry transition can take place at roughly the same temperatures and chemical potentials as the MIT-like model deconfinement transition, meaning that both first order phase transitions can take place at the same time.\\

{\bf Acknowledgments} 

This work is a part of the project INCT-FNA Proc. No. 464898/2014-5. 
D.P.M. and K.D.M. are partially supported by Conselho Nacional de Desenvolvimento
Científico e Tecnológico  (CNPq/Brazil) respectively under grant 301155.2017-8  
and with a doctorate scholarship. C.B. acknowledges a doctorate
scholarship from Coordenação de Aperfeiçoamento de Pessoal do Ensino Superior (Capes/Brazil).

\end{document}